\documentclass[twocolumn]{revtex4}

\usepackage{graphicx}

\usepackage{amsfonts}

\begin{document}

\title{Diversity-induced resonance}
\author{Claudio J. Tessone$^{1,2}$, Claudio R.
Mirasso$^2$, Ra\'{u}l Toral$^{1,2}$ and James D. Gunton$^3$}

\affiliation{1- Instituto Mediterr\'{a}neo de Estudios Avanzados (IMEDEA), CSIC-UIB, Ed. Mateu Orfila, Campus UIB,
E-07122 Palma de Mallorca, Spain \\
2- Departament de F\'{\i}sica, Universitat de les Illes Balears,
E-07122 Palma de Mallorca, Spain\\
3-Department of Physics, Lehigh University, Bethlehem PA 18015, USA
}
\begin{abstract}
We present conclusive evidence showing that different sources of diversity, such as those represented
by quenched disorder or noise, can induce a resonant collective
behavior in an ensemble of coupled bistable or excitable systems.
Our analytical and numerical results show that when such systems are
subjected to an external subthreshold signal, their response is
optimized for an intermediate value of the diversity. These findings
show that intrinsic diversity might have a constructive role and suggest that
natural systems might  profit from their diversity in order to
optimize the response to an external stimulus.
\end{abstract}
\date{\today} 
\maketitle

Noise induced, or stochastic, resonance emerged in the early
eighties as a proposal to explain the periodicity observed in the
Earth ice ages \cite{BSV81,NN81}. The mechanism is such that an
external forcing acting upon a nonlinear system can be conveniently
amplified under the presence of the right amount of noise. This
innovative proposal led many researchers to look for a similar {\it
constructive} role of noise in physical, chemical, biological, and
many other kinds of systems \cite{MBS93,GHJ98,LGN04,WSB04}. While
initially the studies focused on simple, low--dimension, dynamical
systems, more recent work \cite{HDW01,BPT94} has considered the
constructive role of noise in extended systems composed of many
coupled identical units. The assumption of identical units, while
being mathematically convenient, is not very realistic for many of
the applications since it is clear that in some natural systems,
specially in biology, the units composing the ensemble present a
disparity in the values of some characteristic parameters. Among
other consequences, this natural diversity makes each isolated
system respond differently to an external forcing; it is an open
question to investigate the effect that diversity has on the global
response of the collective system.

This problem has received some recent attention. For instance,  Hong \cite{Hong05} analyzes the locking behavior of an
ensemble of coupled oscillators with different internal frequencies subject to a periodic external forcing. He finds that
the quenched disorder helps a small fraction of the oscillators to lock to the external frequency. However, he does not
observe a collective behavior in which the whole ensemble benefits form the diversity in the internal frequencies.
In this Letter we give evidence that the right amount of diversity, in the form of quenched noise, might help an extended system to respond
globally in a more coherent way to an external stimulus.

As in the case of stochastic resonance, we believe that the results reported here are very general. For the sake of
concreteness, however, we have considered two prototypical nonlinear systems: one bistable and another excitable.
In both cases, we show that there is a resonance effect in the global response as a function of the diversity. 

We consider first an ensemble of $N$ globally coupled bistable systems, whose dynamics is given by
\begin{equation}\label{eq:def}
\dot{ x_i } =  x_i -  x_i^3 + a_i + \frac C N \sum_{j=1}^N
\left( x_j- x_i\right) + A \, \sin (\Omega t).
\end{equation}
Here $x_i(t)$, $i=1,\dots,N$ is the position of the $i$-th unit at
time $t$ and $C$ is the  coupling strength. The location and
relative stability of the fixed points of the dynamics of an
isolated unit $i$ are modified by the parameter $a_i$. We assume the
$a_i$'s to take independent values distributed according to a
probability distribution function $g(a)$ that satisfies $\langle a
\rangle= 0$, $\langle a_i \, a_j \rangle = \delta_{ij} \sigma^2$.
$\sigma$ will be referred to as the {\em diversity}. The system is
also subjected to an external periodic forcing, of intensity $A$ and
frequency $\Omega=2\pi/T$.  For simplicity, and to emphasize the
role of the diversity, we neglect
 standard noise effects on those equations.

In order to quantify the response of the system to the external forcing, we introduce  the average position of
the units $X(t) = { 1 \over N }\sum_{i=1}^N x_i(t)$. In the globally coupled case considered here, the coupling amongst
units appears only through this macroscopic quantity:
\begin{equation}\label{eq:reduced}
\dot{ x_i } = C X + (1-C) x_i -  x_i^3 + a_i + A \, \sin (\Omega t).
\end{equation}
This can be put in a potential form $\dot{ x_i }=-\frac{\partial
V_i}{\partial x_i}$ with a suitable time dependent potential
$V_i(x_i;t,a_i,X)$ whose explicit expression can be easily obtained.
By averaging eq.(\ref{eq:reduced}) over the whole population, we
obtain
\begin{equation}\label{eq:final}
\dot X = X - \frac{1}{N} \sum_i x_i^3 + A \, \sin(\Omega \, t).
\end{equation}
Here and henceforth averages with respect to the variables $a_i$ are replaced with averages
with respect to the distribution $g(a)$.
Following reference \cite{desai78}, let us introduce $\delta_i$, such that $x_i = X + \delta_i$.
We additionally introduce $\frac 1 N \sum_i \delta_i^2 = M$. Notice that $M\ge 0$.  Under the assumption of $\delta_i$ being distributed according to an even distribution\cite{comment} we get, using eq.~(\ref{eq:final}),
\begin{equation}\label{eq:A}
\dot X = X \left( 1 - 3 M \right) - X^3 + A \, \sin(\Omega \, t).
\end{equation}
The unforced system is bistable with equilibrium points at $X_{\pm}=\pm \sqrt{1-3M}$. For $\sigma=0$, $M$ vanishes after an initial transient to wash out the effect of the possibly different initial conditions for the $x_i$'s.
 A weak, subthreshold, forcing (namely $A\lesssim 0.3$ for the range of frequencies used in this work), will not suffice
 to have the global variable $X$ jump from one stable point to the other as it will simply make small oscillations
 around one of the equilibrium points. As the diversity increases, $M$ increases with a twofold effect: first,
 the stable points approach each other and, second, the height of the barrier separating them decreases. It might be possible
 that the weak external forcing is now able to overcome the reduced barrier and the global variable $X$ exhibits wide
 oscillations between the two fixed points following the external forcing. If the diversity increases even further,
 leading to $M>1/3$, the barrier disappears, the two fixed points merge at  $X_0=0$ and the global variable makes
 small oscillations around this new fixed point. We then predict a resonance effect for intermediate values of the diversity
 for which the amplitude of the oscillations of $X$ will be maximum.

It should be clear now what the mechanism is leading to the
resonance. In the homogeneous case, when all systems have $a_i=0$,
the subthreshold forcing can not overcome the potential barrier for
any of them. As the diversity increases, there will be a
number of units for which the value of $a_i$ is such that the
forcing is now suprathreshold for them and the barrier can be
overcome in one direction. These units are able, through the
coupling term, to pull the other units and hence produce a
collective, macroscopic, movement following the variation of the
external forcing. For too large diversity, however, some of  the units to be pulled offer too much resistance to follow the external force and this effect can not be overcome by the favorable units.

Before we present the numerical results sustaining this diversity-induced resonance,
let us present a simplified treatment that allows us to reproduce the aforementioned effect.
The main problem is to obtain the variation with time of the second moment $M(t)$. The classical treatment
of reference\cite{desai78} consists in writing down a hierarchy of equations which is truncated under
some Gaussian approximations for the moments. We follow here an alternative approach. We focus on the regime of
{\em slow} forcings where we can assume that the position $x_i(t)$ is locked at the absolute minimum value of
the potential $V_i(x_i)$ and its dynamics simply follows from the movement of that minimum. Given $X(t)$,
this allows us to find $x_i(t;a_i)$ as a function of the parameter $a_i$. Once the $x_i(t)$'s have been computed
in this way, the value of $M(t)$ follows from its definition
$M(t)  = \int da \,g(a)\left[x(t;a)-X(t)\right]^2 $ . This can now be put back in eq.~(\ref{eq:A}).
The whole process can be implemented very efficiently in a numerical integration scheme for $X(t)$.
We quantify the resonance effect by the spectral amplification factor \cite{JH-1989},  $\eta = 4A^{-2} \left| \langle \hbox{e}^{i \, \Omega t} X(t) \rangle \right|^2$, and $\langle \dots\rangle$ denotes a time average.

\begin{figure}
\begin{center}
\includegraphics[width=7cm,angle=-90]{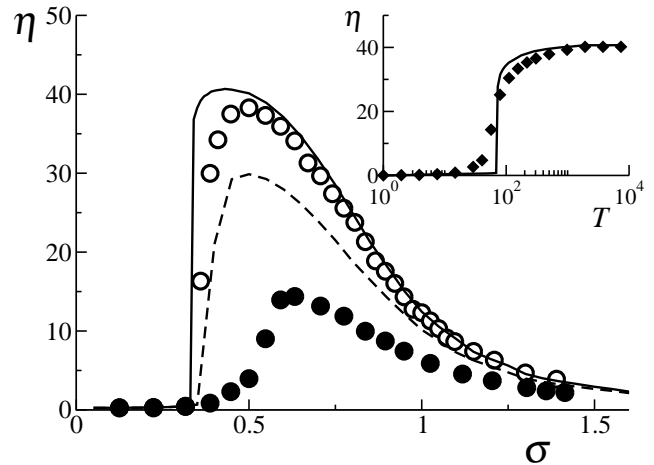}
\caption{\label{fig1}Spectral amplification factor, $\eta$, of the
globally-coupled bistable model, eqs.~(\ref{eq:def}). The values of
the $a_i$'s are drawn from a Gaussian distribution of zero mean and
variance $\sigma^2$. Some system parameters, are: $N=10^3$, $C=1$,
$A = 0.20$. In the main plot we observe that the amplification
factor exhibits a maximum as a function of the diversity both in the
case of a period $T=20$ (black circles) and $T=10^3$ (open circles)
of the external forcing. The inset plots the spectral amplification
factor as a function of the period of the forcing (the diversity is
fixed at $\sigma=0.55$). In both plots, symbols correspond to
numerical simulations and the lines are the corresponding
theoretical predictions of a simplified theory (see the text for
details). }
\end{center}
\end{figure}

\begin{figure}
\begin{center}
\includegraphics[width=7cm,angle=0]{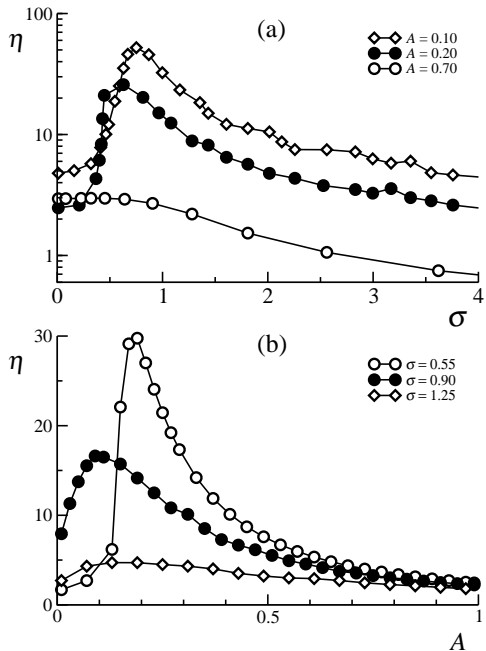}
\caption{\label{fig2}(a) Spectral amplification factor, $\eta$, of the
globally coupled bistable model, eqs.~(\ref{eq:def}) as a function of diversity for
different values of the amplitude $A$ of the forcing. (b) Dependence of $\eta$ on the
forcing amplitude for fixed values of diversity. Both figures use a period $T=50$ and
other parameters as in  fig.\ref{fig1}. Symbols represent the results
coming from a numerical simulation of the system's equations (the solid line is a guide to the eye).
}
\end{center}
\end{figure}

In fig.~\ref{fig1}, we plot the amplification factor $\eta$ versus the diversity $\sigma$, for
different values of the period $T$ of the external forcing for an amplitude $A$ below the
threshold value. As predicted, there is an optimum value of the diversity for maximum
amplification, the main result of this Letter. Notice that our approximate treatment agrees
rather well with the results coming from a direct numerical integration of the original set
of equations (\ref{eq:def}), when the signal is slow.

We now analyze how the system responds to different modulation
periods of the external forcing while the amplitude is kept fixed.
In the inset of fig.~\ref{fig1} we plot the amplification factor as
a function of the period of the external forcing for  fixed
diversity. It can be seen that for large $T$ the
amplification factor reaches a constant value, while  $\eta$
vanishes for small $T$. Both regimes are well described by the
theoretical approximation. For large $T$ the agreement is due to the
validity of our approximate picture of the dynamics in that limit.
For small $T$,  the individual units are not able to follow the fast
external forcing and consequently $\dot x_i \approx 0$ which leads
to the same condition to determine $x_i$ as a function of $a_i$ and
$X$ as in the large $T$ limit.  It is worth mentioning that the
shape of the curve in the inset of fig.~\ref{fig1} differs from what
appears in stochastic resonance where a maximum at intermediate
values of $T$ is observed\cite{GHJ98}. This difference is due to the absence in
the diversity-induced resonance case of a matching between two time
scales which in stochastic resonance are the Kramer's time and the
forcing period.

In fig.~\ref{fig2}(a), we study the effect of the amplitude of the forcing
on the system response. As in
stochastic resonance \cite{GHJ98}, a maximum in the response appears only
for subthreshold forcing and
the height of this maximum increases with decreasing amplitude. However, for suprathreshold forcing (the case $A=0.7$) the
linear regime is recovered and the amplification factor steadily decreases
 with increasing diversity. Fig.~\ref{fig2}(b) shows that the spectral amplification factor has a maximum for a well defined value
 of the amplitude of the external forcing.

We now turn our attention to excitable systems. As a paradigmatic model of interest in many biological
applications, we consider a globally coupled ensemble of excitable units described by the
FitzHugh-Nagumo equations:
\begin{eqnarray}
\epsilon\dot{x_{i}}& = &
x_{i}-\frac{1}{3}x_{i}^3-y_{i}+\frac{C}{N}\sum_{j=1}^N
(x_{j}-x_{i}), \nonumber \\
\label{eq:FN} \dot{y_{i}}& = & x_{i}+a_i + A \sin\left(\Omega \, t
\right).
\end{eqnarray}
The coupling between units is taken into account through the activator variable $x$ with a
coupling strength $C$. Each unit has a parameter $a_i$, representing the diversity, drawn from a
probability distribution $g(a)$ of mean $\langle a_i \rangle =  a$ and correlations
$\langle (a_i-a)(a_j -a)\rangle = \delta_{ij} \sigma^2$. When $|a_i|<1$ system $i$ is in the
oscillatory regime, while for $|a_i| \geq 1$ it is in the excitable one. As in the double well case,
the system is subjected to a periodic forcing of intensity $A$ and frequency $\Omega$ and we do
not consider explicit noise terms. The combined effect of diversity and noise was considered
in reference \cite{ZKH01} in the context of coherence resonance. Specifically, the authors of this
reference found that in the  unforced case, $A=0$, and in the presence of noise, there was a
systematic increase of the coherence factor for increasing inhomogeneity. We focus in this Letter
on the forced case, $A\ne 0$ where we will show a resonance effect with respect to the diversity.

The theoretical analysis follows the lines of the double-well system. With the definitions
$X = \frac 1 N \sum_i x_i$, $Y = \frac 1 N \sum_i y_i$ and $M=\frac 1 N \sum_i (x_i-X)^2$ we arrive at
\begin{eqnarray}
\epsilon\dot X &=& X (1-M) - \frac{X^3}{3} - Y, \nonumber \\
\dot Y &=& X + a + A \sin\left(\Omega \, t\right). \label{eq:FNG}
\end{eqnarray}
We conclude that in this model an increase in the diversity, hence
an increase in $M$, induces a change in the shape of the nullclines
of the dynamics of the global variables. As in the double well
system, in the homogeneous case, $a_i=a$, and $|a|>1$ all units are
in the excitable regime and we consider the case where the weak external forcing is not enough
to overcome their excitability threshold. As the diversity
increases, some units will have their excitability threshold lowered
(they could even become oscillatory) and the forcing
 is now suprathreshold for them. Those units pull the others, so producing the observed collective  behavior.
 The actual description of the collective behavior is somewhat more involved, since $M$ exhibits a periodic
 variation with time and it has a maximum value when the collective variables $X$ and $Y$ are near the fixed point.
 In the limit of large $M$ the nullclines are modified such that the limit cycle disappears altogether.

\begin{figure}
\begin{center}
\includegraphics[width=7cm,angle=0]{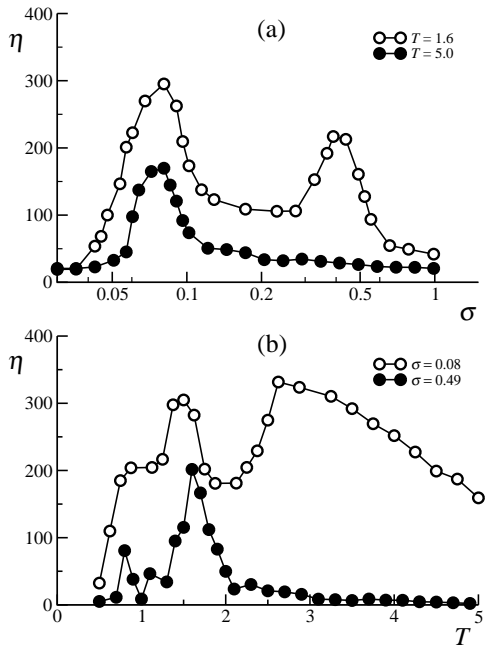}
\caption{Spectral amplification factor, $\eta$, of the
globally-coupled FitzHugh-Nagumo model, eqs.~(\ref{eq:FN}), where the $a_i$'s have been drawn from
a Gaussian distribution of mean $a$ and variance $\sigma^2$.  Some system parameters, are: $N=10^3$,
$\epsilon = 10^{-2}$, $a = 1.12$, $C=1$, $A= 0.05$. (a) Plot of $\eta$ as a function of the diversity $\sigma$
for different periods of the external forcing. (b) Plot of $\eta$ as a function of the period $T$ for
different values of the diversity $\sigma$. In both cases the symbols represent the results coming from
a numerical simulation of the system's equations (the solid line is a guide to the eye).
\label{fig3}}
\end{center}
\end{figure}

In fig.~\ref{fig3}(a) we plot the amplification factor $\eta$ of the
global $X$ variable as a function of the diversity $\sigma$ for
different values of the external time period $T$ and a fixed value
of the  amplitude $A$ close to threshold, where we can observe the
resonance effect. This plot shows some differences with the double
well system studied before, namely the presence of several
resonances at different values of the diversity. We speculate that
this behavior has its origin in the existence of a well defined
refractory time in the dynamics of an isolated unit. Several
resonance maxima can also be observed when plotting the
amplification factor as a function of the period of the forcing for
fixed diversity, see fig.~\ref{fig3}(b). A similar effect has been
also reported for a single FitzHugh-Nagumo system in the presence of
noise and it is known as frequency-dependent stochastic resonance
\cite{Kurths03}.

In conclusion, we have given evidence that diversity, in the form of quenched noise, can enhance and lead to a resonant effect for the response of an
extended system to an external periodic forcing. The evidence has been given for
two prototype systems, paradigmatic
of bistable and excitable behavior and, hence, we believe that the same resonance
will appear in other more
complicated systems. The mechanism of the phenomenon is particularly simple: at
 a given time a fraction of the units are
able to respond to the external forcing; those units, through the
coupling terms, are able to pull the others into the direction of
the force. For too large diversity, the favorable units can not overcome the effect of the adverse ones.
This resonance mechanism is very general and it could appear in many fields.

A final remark is relevant here. Note that in equation
(\ref{eq:A}) for the global variable $X$  the effect of the
diversity appears only through the variable $M$ measuring the
dispersion in the behavior of the dynamics of the individual units
$x_i$. Therefore, the existence of a resonance effect for the
optimal amplification of weak signals {\sl does not depend on the
source of the disorder}. The same effect could also be obtained in
the presence of disorder induced by noise (stochastic resonance), by
a non-regular network of connectivities, inhibitory couplings, etc.

The idea that different sources of diversity can produce a resonant effect leads
us to  speculate that the amount of
diversity present in some biological systems has an important function. Diversity
could have been evolutionary tuned in order to enhance the detection of weak signals. Whether
natural systems have taken advantage or not from this diversity related effect is a question that,
as in the particular case of stochastic resonance, has not yet a clear answer.

\acknowledgments

We acknowledge financial support by the Ministerio de Educaci\'on y
Ciencia (Spain) and FEDER projects FIS2004-5073, FIS2004-953, the EU
NoE BioSim LSHB-CT-2004-005137, NSF Grant DMR-0302598 and the G.
Harold and Leila Y. Mathers Charitable Foundation. CRM acknowledges the hospitality and support of the Lehigh university.


\begin{thebibliography}{14}
\bibitem[{com()}]{comment}
\bibinfo{note}{Alternatively, one could simply neglect the third moment of the
  distribution of $\delta_i$, an assumption valid for small $\delta_i$.}
\end{thebibliography}
\end{document}